\pdfoutput=1
\documentclass[english]{article}
\usepackage{geometry}
\usepackage[utf8]{inputenc}				
\usepackage[T1]{fontenc}    
\usepackage{hyperref}       
\usepackage{url}            
\usepackage{booktabs}       
\usepackage{amsfonts}       
\usepackage{nicefrac}       
\usepackage{microtype}      
\usepackage{lmodern} 
\usepackage{amsmath}
\usepackage{authblk}
\usepackage[authoryear,sort&compress]{natbib}

\usepackage{varwidth}
\usepackage[format=hang]{caption,subfig} 
\usepackage{enumitem}
\usepackage{array}   
\newcolumntype{R}{>{$}r<{$}} 
\usepackage{wrapfig}

\usepackage{graphicx}
\graphicspath{ {./figures/} }

\title{Causal effect of the infield shift in the MLB}
\date{}

\author[1]{Sonia Markes\footnote{sonia.markes@mail.utoronto.ca}}
\author[1]{Linbo Wang\footnote{linbo.wang@utoronto.ca}}
\author[1]{Jessica Gronsbell\footnote{j.gronsbell@utoronto.ca}}
\author[2]{Katherine Evans\footnote{kathy.l.evans@gmail.com}} 

\affil[1]{University of Toronto, Department of Statistical Sciences, Toronto, ON, Canada}
\affil[2]{Washington, DC, USA}

\begin{document}

\maketitle

\abstract{The infield shift has been increasingly used as a defensive strategy in baseball in recent years. Along with the upward trend in its usage, the notoriety of the shift has grown, as it is believed to be responsible for the recent decline in offence. In the 2023 season, Major League Baseball (MLB) implemented a rule change prohibiting the infield shift. However, there has been no systematic analysis of the effectiveness of infield shift to determine if it is a cause of the cooling in offence. We used publicly available data on MLB from 2015-2022 to evaluate the causal effect of the infield shift on the expected runs scored. We employed three methods for drawing causal conclusions from observational data---nearest neighbour matching, inverse probability of treatment weighting, and instrumental variable analysis---and evaluated the causal effect in subgroups defined by batter-handedness. The results of all methods showed the shift is effective at preventing runs, but primarily for left-handed batters.}


\section{Introduction}
\begin{figure}[!ht]
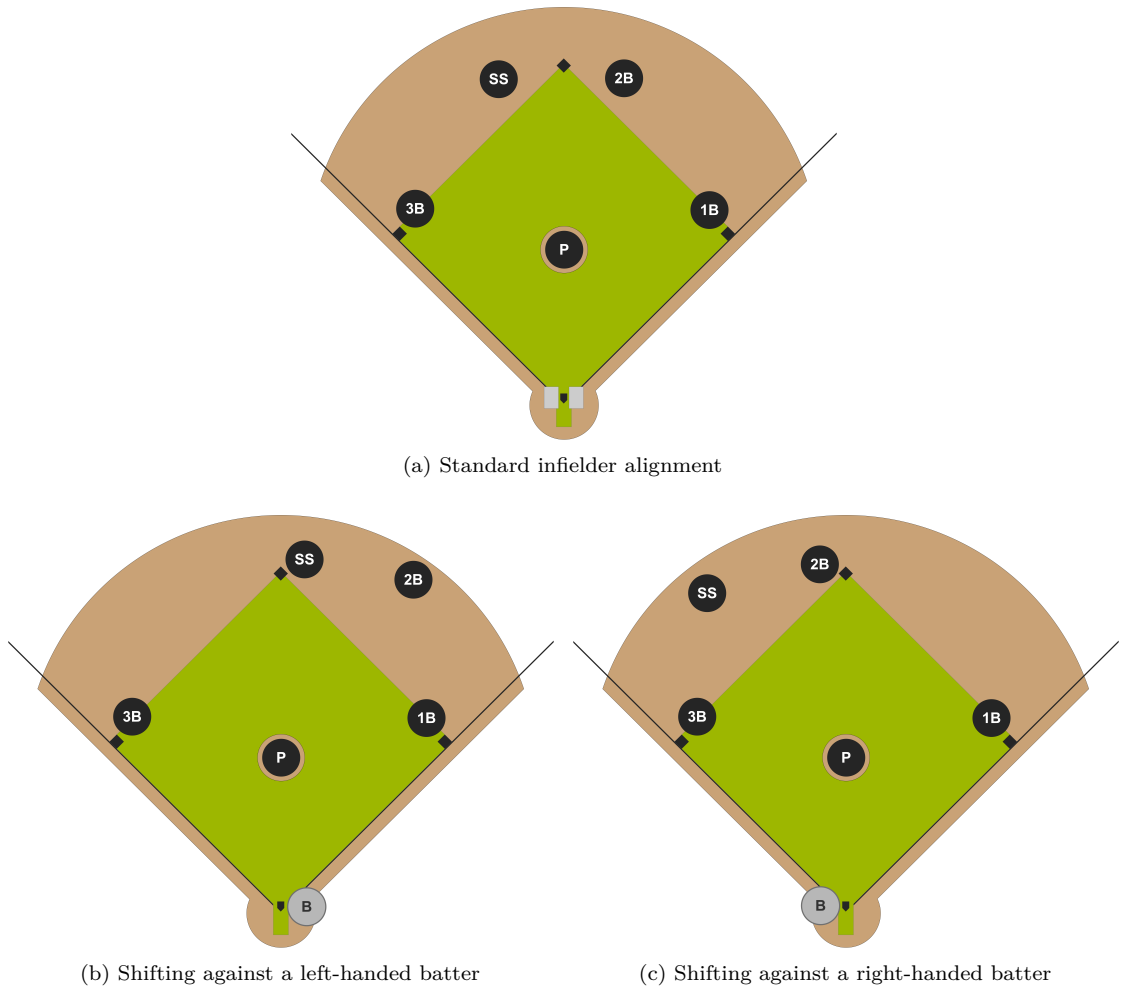

\centering
\subfloat[Standard infielder alignment]{
    \includegraphics[width = .45\linewidth]{infieldalign_standard.pdf}
    \label{fig:alignstandard}} \\
\subfloat[Shifting against a left-handed batter]{
    \includegraphics[width = .45\linewidth]{infieldalign_LHshift.pdf}
    \label{fig:alignLH}}
\subfloat[Shifting against a right-handed batter]{
    \includegraphics[width = .45\linewidth]{infieldalign_RHshift.pdf}
    \label{fig:alignRH}}
\caption{Two players stand on either side of the line defined by home plate and second base in a standard infielder alignment \protect\subref{fig:alignstandard}, whereas the infield shift is defined by three of the four infielders standing on one side of the field. Which side has three infielders depends on the side of the plate that the batter stands on \protect\subref{fig:alignLH}-\protect\subref{fig:alignRH}.}
\label{fig:alignments}
\end{figure}

Major League Baseball (MLB) teams have developed sophisticated models to determine the optimal positions for infielders to be most effective in a given game situation. While few studies on optimal placements are published (see \citet{BouzarthEtAl2021}), analyses of infielder positioning have identified relevant factors to the decision models, such as tendencies of the pitcher \citep{Choi2021} or of the batter \citep{Gerli2019}. In particular, batters often have a tendency to hit towards the side of the field that aligns with the side of the plate they hit from, called the pull side, with some batters showing stronger tendencies than others. To take advantage of this tendency, teams sometimes move three infielders to the batter's pull side of the field, as shown in Figure \ref{fig:alignments}, a defensive strategy known as the \emph{infield shift}. However, using the infield shift carries a risk as it leaves the opposite side of the field exposed, which can be a liability if a hit goes towards the opposite field and creates running opportunities for the batting team, especially if they have players on base. Despite this, the use of the infield shift has increased in recent years---the rate of shifted plate appearances has tripled from 2015 to 2022---suggesting that teams believe the benefits outweigh these risks. 

Concerns that the infield shift may have contributed to decreases in offensive production led the MLB to implement rule changes for the 2023 season, which eliminated the infield shift from the game \citep{MLB23RuleChange:Castrovince2023}. However, an \emph{association} between using the infield shift and decreased offensive numbers does not imply that the infield shift is a \emph{cause} of the low offence. Furthermore, it has not been established that the shift has been causally effective against the targetted batters. Some research suggests that batters have adapted to it \citep{Sawchik2019}. Has the infield shift caused lower offensive numbers when it has been utilized? This paper aims to address this question by estimating the causal effectiveness of the infield shift as a defensive strategy.


Causality is most reliably established with randomized control trials, but only observational data is available for baseball. A range of methodologies for asking causal questions using observational data have been developed in statistics, computer science, and econometrics \citep{WhatIf:Hernan+Robins2020,Causality:Pearl2009,MostlyHarmlessEconometrics:AngristEtAl2008}. These methods have seldom been used in sabermetrics (an exception is \citet{Vock+Vock2018}), and, to the best of our knowledge, formal causal inference methods have not been used to assess the infield shift. Previous investigations of how the shift works have been association studies \citep{Gerli2019,Sawchik2019,Eassom2018,Lindbergh2020}, focused on different questions \citep{LeDoux2018,BouzarthEtAl2021,Choi2021}, or have used ad-hoc methods (e.g. using only specific subsets, not accounting for confounding) to consider causal questions \citep{Carleton2018,Tango2020,Eassom2019,Eassom2019hand}. Our causal analysis builds on insights from these works to assess the shift. 

In causal inference, different methods require different assumptions \citep{WhatIf:Hernan+Robins2020,BaiocchiEtAl2014}, each carrying distinct risks of leading to incorrect conclusions. We utilize multiple methods to assess the effectiveness of the infield shift for plate appearance when it has been deployed, minimizing reliance on a single set of untestable assumptions. In particular, we use an instrumental variables analysis, a method which, to the best of our knowledge, is novel to sabermetrics. In this paper, we evaluate the causal effect of the infield shift on the expected runs scored using publicly available observational data from Statcast. The question we address is: Has the infield shift been effective at causing suppression of expected runs when the infield shift has been deployed? 

The paper is organized as follows: Section \ref{sec:framework} situates the problem of the infield shift within a causal inference framework and Section \ref{sec:data} details the dataset used. Section \ref{sec:methods} details the statistical analysis and Section \ref{sec:results} lays out the results of the three methods used: matching, inverse probability of treatment weighting (IPTW), and instrumental variable analysis. Section \ref{sec:discussion} is a discussion of the results of the different approaches. While all methods indicate that the shift is effective in preventing runs against left-handed batters, the conclusion regarding right-handed batters depends on the assumptions one is willing to make.

\section{Framework} \label{sec:framework}
A scientific experiment aims to evaluate the effect of changing a quantity of interest, with all else held equal; for example, two groups---treated and untreated---that are the same in every way, except for the quantity of interest. Random assignment to treatment from a sample is the ideal way to achieve this. However, observational data does not have the advantages of a randomized experimental design. Causal inference aims to draw causal conclusions from observational data by emulating a randomized control trial. \citep{WhatIf:Hernan+Robins2020}

We were interested in studying the effects of infielder alignment shifts compared to not shifting. Ideally, we would run an experiment in which the shift is randomly applied, making every situation equally likely to be shifted. This includes every combination of batter, pitcher, infielders, game context, and fielding strategy. However, fielding teams do not implement shifts at random. Instead, the decision is guided by attributes of the batter, pitcher, infielders, game context, and fielding team strategy. There may also be unrecorded or unmeasured factors, such as player injuries.

\begin{figure}[!ht]
\centering
\includegraphics[]{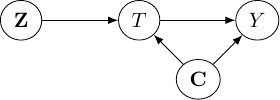}
\caption{An example DAG where $T$ represents the treatment variable and Y is the outcome variable. The covariates included in this DAG are confounders, $\mathbf{C}$, and instruments, $\mathbf{Z}$.}
\label{fig:DAGex}
\end{figure}

To approximate a randomized control trial, our approaches rely on other observed covariates, in particular, confounders and instruments. Confounders are variables that affect both treatment and outcome, whereas instruments only affect the outcome through the treatment and are assumed to be independent of unmeasured confounding variables. Models for causal relationships can be represented by directed acyclic graphs (DAGs), as shown in Figure \ref{fig:DAGex}.

\begin{figure}[!ht]
\centering
\includegraphics[]{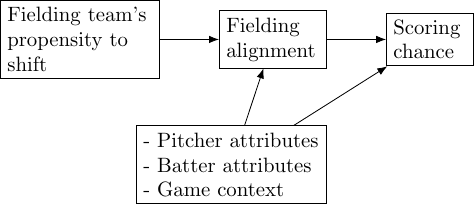}
\caption{The high-level DAG used for this analysis.}
\label{fig:DAG}
\end{figure}

Variables were chosen and classified in our model based on a combination of insights from other research and input from domain experts in baseball (see Acknowledgement). The fielding alignment for each plate appearance takes the role of a so-called ``treatment'' variable and is represented as a dichotomous variable, $T \in \left\{ 0 , 1\right\}$ indicating whether infielders positioned in standard alignment or shifted alignment, respectively. The efficacy of a defensive effort could be evaluated based on the change in the offensive team's potential to score runs or the actual change in the number of runs scored. An effective defensive strategy suppresses scoring or changes the scoring chances for the worse. The traits of the batter and the pitcher, as well as the context of the game, were looked at as confounders. For example, a batter's average launch angle was considered as it impacts the trajectory of the ball (and therefore, the potential to score runs). Fly balls were found to be increasing with the number of balls in play as the use of the infield shift increased \citep{Sawchik2019}, suggesting a batter's launch angle could confound the effect of the shift on scoring potential. The fielding team's propensity for shifting was considered as an instrument, as teams with high shift rates are not necessarily the most winning teams (this choice is discussed in more detail in Section \ref{sec:IVs}). A representation of this model is summarized in the high-level DAG shown in Figure \ref{fig:DAG}. This was used to guide the choices made with the data and analysis.

Controlling for confounders is an intuitive approach to causal inference from observational data. A straightforward way of doing this would be to select every observation with the same combination of  \{batter, pitcher, and game context\} and compare outcomes of shifted plate appearances with plate appearances that were not shifted (e.g. \citet{Tango2020}). However, each particular combination may not appear very often, even over multiple seasons. Thus, this method would not be generalizable and would be vulnerable to bias due to post-selection. Instead of relying on the identity of a batter, pitcher, or fielding team, we made use of the extensive data collected about each game, characterizing players by attributes and measurements of their skill and tendencies. This approach has the advantage of allowing more data to be used and results to be applied more broadly.

A batter's handedness plays a special role. Bases are run counter-clockwise, regardless of which side of the plate the batter hits from or which side of the field they tend to hit toward. The direction of base running induces an asymmetry into the fabric of the game, so leaving one side of the field vulnerable has different risks depending on the batter's handedness. Previous analyses of the infield shift have found different effects depending on the handedness of the batter; \citet{Gerli2019} found that the infield shift was associated with lower batting averages on ground balls, even more so for left-handed batters. As such, we consider batter-handedness as an effect modifier, and we carried out our analysis on subgroups defined by batter-handedness.

\section{Data} \label{sec:data}
\subsection{Source}
We sourced the dataset for this analysis from \href{http://MLB.com}{MLB.com}'s publicly available Statcast database, accessed using a scraping function \citep{R-baseballr} to pull data from eight seasons, 2015-2022 inclusive. The Statcast database has pitch-level observations, with identifiers for the batter, the pitcher, and their teams. Each observation includes information about pre-pitch game context, fielding alignment, various measurements of the pitch and of any hits, and the resulting impact on the score. Using this information, we defined a treatment variable representing shifted infielder alignment and chose a measure of the outcome.

As a treatment variable, we used a binary definition of the infield shift---alignment of infielders is shifted or not shifted---as identified in the Statcast database by the infield fielding alignment data field (\texttt{if\_fielding\_alignment}). Infield fielding alignment has three possible values in the Statcast database: standard, shifted, or strategic \citep{def_shifts}. Standard alignment means there are two infielders on either side of the second base and each fielder is within a specific range of home plate, depending on his position. Shifted alignments indicate that three or more infielders are positioned on one side of the second base. If the infielders are in a configuration that is neither standard nor shifted, then it is categorized as strategic. While the publicly available Statcast data treats fielding alignment as categorical, MLB teams have more precise data that includes exact infielder positioning. As a consequence of having continuous infielder position data, fielding teams consider a continuous set of options when positioning their infielders. Although the process of generating fielding alignments creates positions in a continuous space, the public data, and therefore the public discourse, treats the shift as a discrete variable. Since we were concerned with the question of the shift as compared to all other alignments, we combined the standard and strategic observations into a single category. 

As an outcome variable, we used the \emph{change in run expectancy}, given as a field in the Statcast database called \texttt{delta\_run\_exp}. \emph{Run expectancy}, $RE$, is an empirical average of the runs scored in the remainder of the inning, given the pre-pitch game state. Here, the game state is characterized by what is termed the base-out state, the locations of any runners on base and the number of outs, and by the count, that is, the number of balls and strikes. A change in run expectancy, $\Delta RE$, captures the change in the scoring potential resulting from a change in the game state while also accounting for any actual change in the score. More precisely, $\Delta RE$ is calculated as the difference in run expectancy between the final game state and the initial game state, plus any change in the score:

$$
\Delta RE = RE_{final} - RE_{initial} + \Delta score
$$

\noindent
where $RE{state}$ for each configuration of the game state---termed a \emph{run expectancy matrix}---is calculated each year \citep{Tango2018, RE24:Weinberg2014}. The run expectancy matrix can change significantly from year to year, and thus, game year will be an important variable to control for throughout this analysis.

This definition points to a key feature of the change in run expectancy; it captures how the value of the play depends on the context in terms of the year and the initial game state. For example, in 2018, the change in run expectancy of the strikeout is -0.24 runs if the bases are empty, but if the bases are loaded with one out, a strikeout carries a change in run expectancy of -0.81 runs (according to the run expectancy matrix by \citet{Tango2018}). Since the risks of the infield shift are also context-dependent, the change in run expectancy is a good outcome variable to assess the effectiveness of the shift. The use of the shift has also changed from season to season, so we considered each season separately for this analysis. However, changes in run expectancy are often negligible at the pitch level. In fact, the pitch-level change in run expectancy is only non-zero if a runner steals a base or it is the last pitch of a plate appearance. To make each observation more comparable and to ensure a noticeable effect size, we chose to aggregate the data to the level of a plate appearance.

\subsection{Transformation}
We transformed the data from pitch level to plate appearance level. To ensure well-defined treatment and outcome variables, any plate appearances missing a field alignment or a change in run expectancy were excluded. We define a plate appearance to be shifted if the infield fielding alignment is shifted for any pitch of the plate appearance. Calculating the change in run expectancy for a plate appearance involves summing the change in run expectancy across pitches, which is how we obtained the outcome variable. Change in run expectancy for a plate appearance does not depend on the count since the count only changes between pitches; the base-out state encapsulates the plate appearance-level game state. For fidelity, we excluded plate appearances that had missing pitches, as well as any plate appearance where the same batter or the same pitcher did not take all pitches.

The batter attributes we considered for this analysis were handedness, hitting tendencies, skill, and results, as well as, how they have been shifted against previously. Given the importance of batter handedness and spray angle tendencies for an analysis of the infield shift, we excluded batters who hit switch or who do not have any hit trajectory information over a season. The handedness of the batter is represented in the Statcast database by the field \texttt{stand}, indicating the side of the plate the batter stood on. To represent hitting tendencies, we used data fields for launch speed (\texttt{launch\_speed}), vertical launch angle (\texttt{launch\_angle}), and spray angle (calculated from hit coordinates, \texttt{hc\_x} and \texttt{hc\_y} \citep{Petti2020}, with sign convention set relative to batter handedness so that a negative spray angle indicates a pull hit). Specifically, we used summary statistics of these hit measurements (mean and standard deviation, cumulative to the batter's most recent plate appearance). Similar summary statistics were calculated to represent skill and results; weighted on-base average (wOBA) and batting average on balls in play (BABIP), and a batter's number of walks and strikeouts as a proportion of plate appearances can be summarized from similarly named fields in the Statcast database. 




\begin{figure}[!hb]
\centering
\includegraphics[]{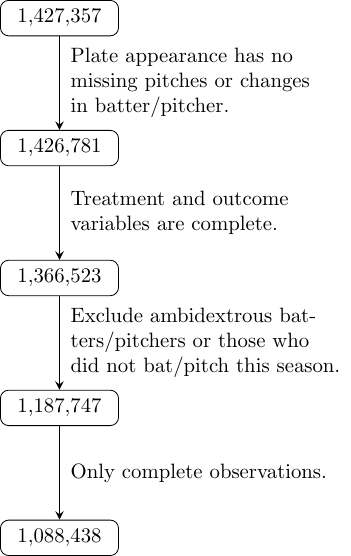}
\caption{Number of plate appearance observations in the dataset after applying each set of requirements.}
\label{fig:counts}
\end{figure}

Pitching is the opposite side of hitting. Accordingly, the pitcher attributes we considered are similar to the batter attributes considered: handedness, pitching tendencies, skill, and results. We excluded pitchers who are ambidextrous or who do not have any pitch trajectory information over a season. As with missing hit trajectory data, it is possible that this excludes observations which are missing not at random. Pitcher handedness is captured in the Statcast data field \texttt{p\_throws}. The attributes of pitching tendencies we used were summary statistics of the speed and spin on release (\texttt{release\_speed} and \texttt{release\_spin\_rate}) and horizontal and vertical position when the ball crosses the plate (\texttt{plate\_x} and \texttt{plate\_z}). Each observation in our dataset has the cumulative mean and standard error of these attributes up to and including the last pitch a pitcher threw before the current plate appearance. These summary statistics were calculated up to the pitch level, so the statistics are weighted by the number of pitches in a plate appearance. Pitcher skill was represented by similar summary statistics of the wOBA and BABIP attributed to a pitcher. 

Summary statistics represented the fielding team's shifting tendencies. For each observation, we calculated the fielding team's shift rate (which we refer to as the fielding team's propensity to shift) and the standard error of that rate on plate appearances up to the start of the current plate appearance. For this analysis, these were the only fielding team attributes we considered.

A full list of the variables, as well as summary statistics or counts, can be found in the Supplementary material.

Observations with missing data were removed from the dataset. We chose to remove these incomplete observations because many of them were created by the lagged fields where imputing values would require stronger assumptions. Removing these observations means that plate appearances were excluded if, from the start of the season up to the start of the plate appearance, the batter had no hits with trajectory information, the pitcher had no pitch information captured, or either the batter or pitcher had made or pitched fewer than 10 plate appearances. Figure \ref{fig:counts} summarizes the impact of each set of exclusions on the number of plate appearances in the analyzed dataset. The biggest decrease in counts came from excluding batters without hit trajectory data or pitchers without pitch trajectory data in a season. Missing trajectory data may be associated with limitations of the measuring instruments, and even with particular ballparks \citep{Arthur2016}, and so, we cannot assume that the data is missing at random. The conclusions of this analysis will apply specifically to batters whose hit trajectory data could be captured over a season and pitchers whose pitch trajectory data could be captured over a season. We return to this point in the Discussion in Section \ref{sec:discussion}.

\section{Methods}\label{sec:methods}

\begin{figure}[!ht]
\centering
\includegraphics[width=0.6\linewidth]{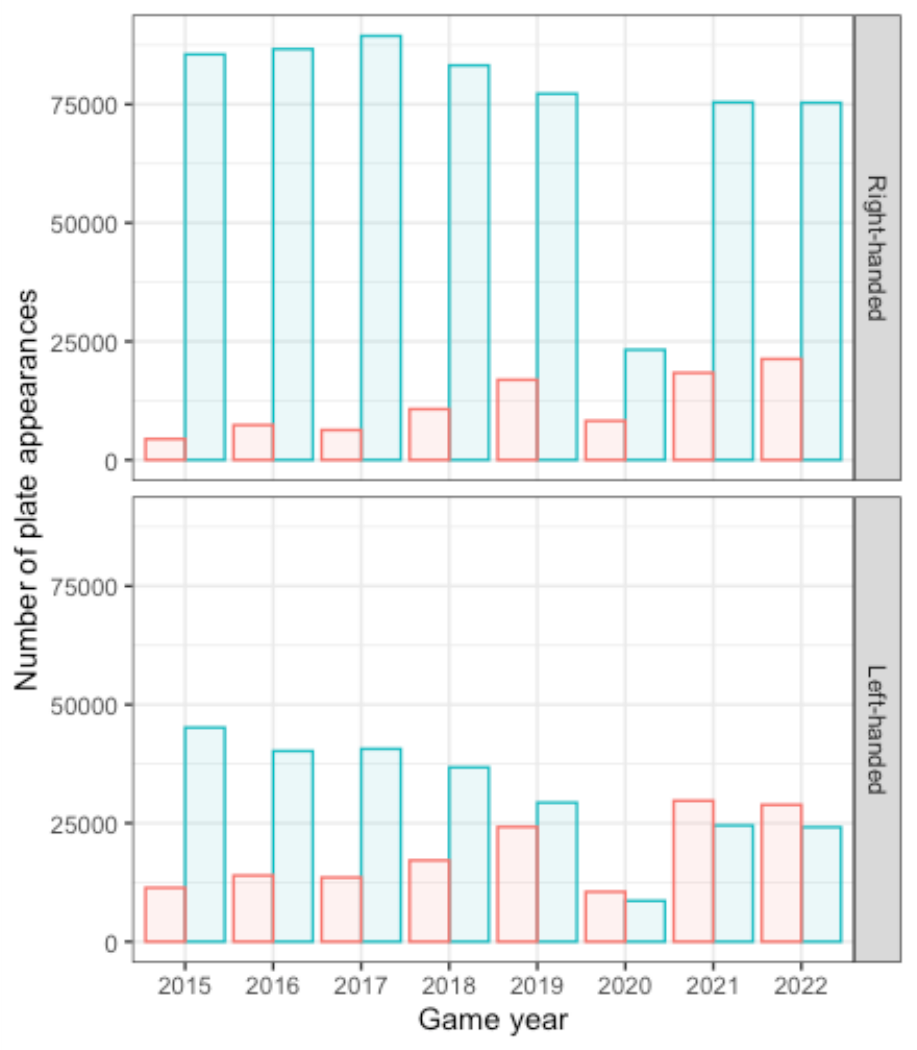}
\caption{Shifted observations (red) are outnumbered by non-shifted observations (blue) in the dataset for all years, except for 2020-2022 for left-handed batters.}
\label{fig:countsTC}
\end{figure}

In this analysis, we are focusing on whether the shift was effective when it was implemented. Our question pertains to the impact on shifted plate appearances, and therefore, our choice of causal estimand aligns with this. We estimated the effect of treatment on the treated (ETT), which is defined as
\begin{equation}
    \mathbb{E}\left[ Y^{t=1} - Y^{t=0} \big| T=1\right]
\end{equation}
where $Y^t$ is the counterfactual outcome variable representing the change in expected runs had the fielding alignment been $t$, and $T$ is the treatment variable representing the fielding alignment, where $T=1$ represents a shifted infield alignment. Our estimates provide an idea of whether the infield shift is an effective strategy for those plate appearances where it was implemented. Note that even with the increasing popularity of alternative strategies, most plate appearances are still defended with a standard fielding alignment, as shown in Figure \ref{fig:countsTC}, so this result does not apply to most plate appearances. See Section \ref{sec:discussion} for further comment. 

The confounders we consider in this model are attributes of batters and pitchers detailed in the previous section, such as handedness, skill, discipline, and tendencies of their hit or throw trajectories, as well as the year. The instrument considered here is the fielding team's propensity to shift across the season to date. In all methods, batter-handedness will also be considered as an effect modifier; that is, we do the entire analysis with subgroups defined by batter-handedness as we want to check if the ETT is different depending on which side of the plate the batter hits from.

    


We employed two approaches to estimate the ETT: balancing methods and instrumental variable (IV) methods. Balancing methods seek to adjust for \textit{measured} confounding, whereas IV methods control for \textit{unmeasured} confounding. Although every method of causal inference requires assumptions for identifiability, these two approaches differ in the set of assumptions they require. At a high level, balancing methods assume all variables that confound the treatment and the outcome are measured and that there is no latent common cause. In contrast, IV methods allow for the possibility of latent common causes of the treatment and the outcome but require assumptions about the instrument's validity. By utilizing both of these methods, we can compare estimates under each set of assumptions without committing to one set of untestable assumptions over another.

One assumption that is common among the approaches we use is known as the \emph{stable-unit-treatment-value assumption (SUTVA)} \citep{Stuart2010, BaiocchiEtAl2014}: There is only one version of the treatment, and the treatment of one subject does not influence the outcome of another. Formally, given $\mathbf{z}=\{z_k\}_{k=1}^N$ and $\mathbf{z'}=\{z'_k\}_{k=1}^N$, if the $i^{th}$ observation is assigned to IV value $z$, that is $z_i=z'_i=z$, then $T_i^{\mathbf{z}}=T_i^{\mathbf{z'}} \equiv T_i^z$. Given also $\mathbf{t}=\{t_k\}_{k=1}^N$, and $\mathbf{t'}=\{t'_k\}_{k=1}^N$, if the $i^{th}$ subject is assigned to IV value $z$ and treatment value $t$, that is, $z_i=z'_i=z$ and $t_i=t'_i=t$, then $Y_i^{\mathbf{z},\mathbf{t}} = Y_i^{\mathbf{z'},\mathbf{t}} = Y_i^{\mathbf{z},\mathbf{t'}} = Y_i^{\mathbf{z'},\mathbf{t'}} \equiv Y_i^{z,t}$. It is reasonable to assume that a shifted plate appearance does not influence the outcome of another batter's plate appearance beyond the confounders that are controlled for. However, the assumption is likely violated by our data by having multiple versions of a treatment since there are theoretically an infinite number of locations for infielders that would satisfy the definition of an infield shift used by Statcast. Since public data does not include the exact coordinates of fielders, we assume the way the infield shift is applied is reasonably similarly each time.

The balancing methods we use here---matching and IPTW---rely on the following assumptions for identifiability \citep{WhatIf:Hernan+Robins2020, Stuart2010}, given a set of confounders $\mathbf{C}$ with estimated propensity score, $\pi(\mathbf{C}) = \Pr (T=1 | \mathbf{C})$ :

\begin{enumerate}
  \item \emph{Partial exchangeability}: Within the levels of the measured confounders, the treatment group would have had the same average outcomes as the control group had they remained untreated, $Y^{t=0} \perp T \: \vert \: \pi(\mathbf{C})$.
  
  \item \emph{Positivity}: Within all levels of the measured confounders, the probability of treatment is non-zero, requiring that $\mathbb{P}[T=t\vert \pi(\mathbf{C})=s]>0$ for all $s$ such that $\mathbb{P}[\pi(\mathbf{C})=s]\neq0$.
  
\end{enumerate}

The IV method builds on potential outcomes $Y^{z,t}$ under IV assignment $Z=z$ and treatment assignment $T=t$ and uses potential treatments $T^z$ under IV assignment $Z=z$. Identification of an IV estimate requires that we make the following assumptions for the ETT \citep{Hernan+Robins2006, BaiocchiEtAl2014}:

\begin{enumerate}
  \item \emph{Relevance}: An IV is associated with treatment, that is, $Z \not\!\perp\!\!\!\perp T \vert \mathbf{X}$.

  \item \emph{Effective random assignment}: An IV is independent of unmeasured confounders, conditional on a set of covariates $\mathbf{X}$, such that 
    \begin{equation} \label{eq:randassign}
        Z \perp \left( \{ T^z \}_{z\in Z}, \{Y^{z,t}\}_{z\in Z, \, t\in T} \right) \; | \; \mathbf{X}
    \end{equation}

  \item \emph{Exclusion restriction}: An IV only affects the outcome through the treatment variable, $Y^{z,t}=Y^{z',t} \equiv Y^t$.

  \item \emph{No effect modification by an IV}, or \emph{no current treatment value interaction}:
\begin{equation} \label{eq:noeffectmod}
    \mathbb{E}[Y^{t=1}-Y^{t=0}|T=t, Z=z, \mathbf{X}] 
            = \mathbb{E}[Y^{t=1}-Y^{t=0}|T=t, Z=z', \mathbf{X}] \quad \forall \, t
\end{equation}

\end{enumerate}

The 4th IV assumption is needed specifically for point identification of ETT (a different assumption would be needed if we were interested in the overall average treatment effect). Also, note that the covariates needed to satisfy the IV assumptions, $\mathbf{X}$, are not mutually exclusive of $\mathbf{C}$. $\mathbf{X}$ includes any variables that represent a possible confounder between the IV and the outcome. A fielding team's propensity to shift and expected runs may be confounded by a team's fielding ability or their pitcher's tendencies. A team's fielding ability can be at least partially accounted for by a pitcher's BABIP statistic \citep{BABIP:FanGraphs}, and the pitcher's tendencies are accounted for by the pitcher's attributes.

If we were to assume that we had controlled for all possible confounding, balancing methods, such as matching or IPTW, would be appropriate. If we assume there are unmeasured confounders, but we have a suitable instrument, then the IV approach can be taken. We do not claim that either set of assumptions is more reasonable, but rather, we present both balancing methods and IV estimation. In the following subsections, we explain how we utilized matching, IPTW, and IVs, and evaluate the suitability of making these assumptions given our dataset. 

\subsection{Matching} \label{sec:matching}
 To construct a matched design, we used linear propensity score as a distance measure to implement $k:1$ nearest neighbour matching \citep{R-MatchIt, HoEtAl2007}. Nearest neighbour matching methods are appropriate for estimating the ETT because controls are matched to treated observations, and those that are not matched are discarded. In our case, there are many more observations of standard fielding alignment, or `controls', than there are of shifted fielding alignment, or `treated'. We chose the tuning parameters to include as many relevant controls as possible without excluding any treated observations or sacrificing covariate balance. The tuning parameters considered were the ratio of controls to treated observations, a caliper on the propensity score distance, and the maximum number of times a control observation could be re-used. Some re-use of controls was necessary to include all treated observations. Although the impact is limited by capping re-use, matching with replacement lowers the precision of the estimate \citep{Austin2013}. Increasing the ratio of control to treated observations can offset this loss in precision by including more controls in the matched set. Using a caliper ensures that controls which are not relevant, as determined by propensity score distance, are not included. We required an exact match on game year and carried out the matching separately on subgroups defined by batter handedness.

\subsubsection*{Choosing tuning parameters}
\begin{figure}[!ht]
\centering
\includegraphics[width=0.7\linewidth]{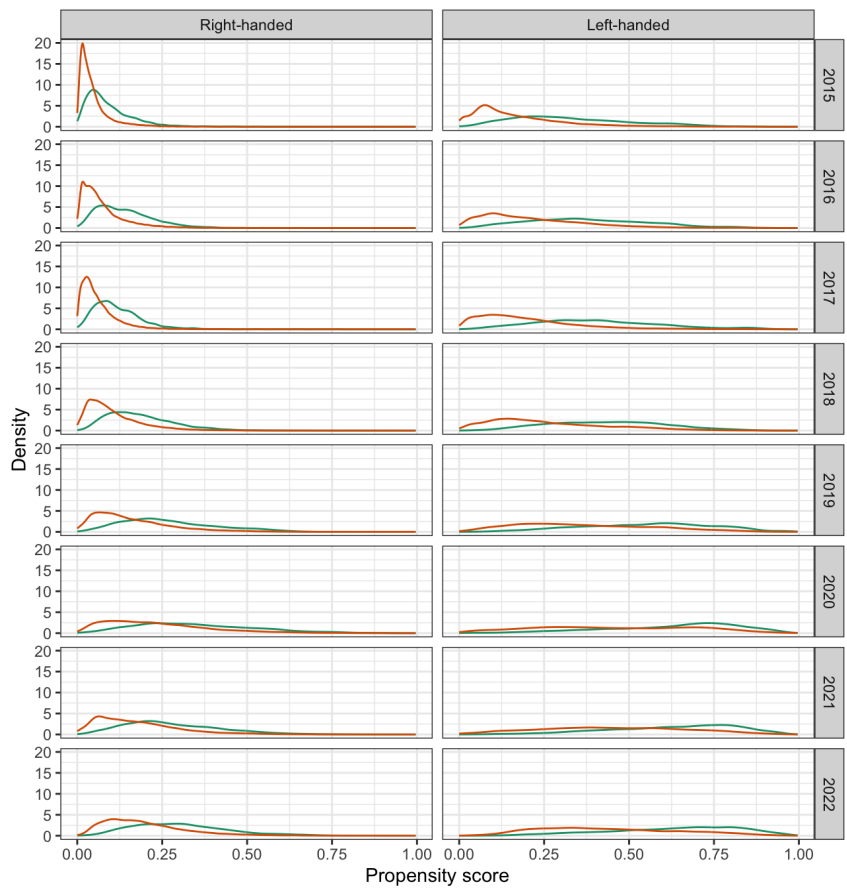}
\caption{The distribution of the propensity scores by year before matching, comparing shifted plate appearances (green) and standard fielding alignment plate appearances (red).}
\label{fig:match1}
\end{figure}

In the original dataset, game year and batter variables differ between shifted and non-shifted plate appearances for both right-handed and left-handed batters. However, for right-handed batters, there was a greater difference in the variance of the propensity scores between treated and control as compared to that of left-handed batters. Figure \ref{fig:match1} shows the distribution of propensity scores is more heavily right-skewed for standard fielding alignment as compared to shifted fielding alignment, particularly for right-handed batters. This difference between shifted and standard fielding alignment propensity scores has been less pronounced in recent years, as might be expected with the steep increase in shift rates across the MLB.

In \citet{Stuart2010}, it is recommended to use a caliper of 0.2-0.25 standard deviations, or smaller if the variance in the treatment group is much larger than the variance in the control group. Even with the increase in popularity of the shift over the past few years, standard fielding alignments have still been used for the majority of the plate appearance, meaning that the data set has more untreated observations than treated. We considered various ratios of untreated to treated observations (1:1, 2:1, 3:1) since matching multiple controls to each treated observation can increase efficiency. However, not all treated observations could be uniquely matched to an untreated observation within a propensity score of 0.25 standard deviations. This discrepancy was particularly pronounced for left-handed batters, excluding 29,788 of 149,288 treated observations, or approximately 20\%. Thus, we allowed for replacement but capped the number of times an untreated observation could be re-used to between 4-6 times and, to avoid poor matches, we focused on caliper values between 0.15-0.2 standard deviations. The Supplementary Material includes a comparison of different matched sets. The matched set chosen for analysis is the set with a 3:1 ratio, a caliper of 0.15, and the re-use maximum set at 5.

\subsubsection*{Balance in the matched set}
\begin{figure}[!ht]
\centering
\includegraphics[width=0.8\linewidth]{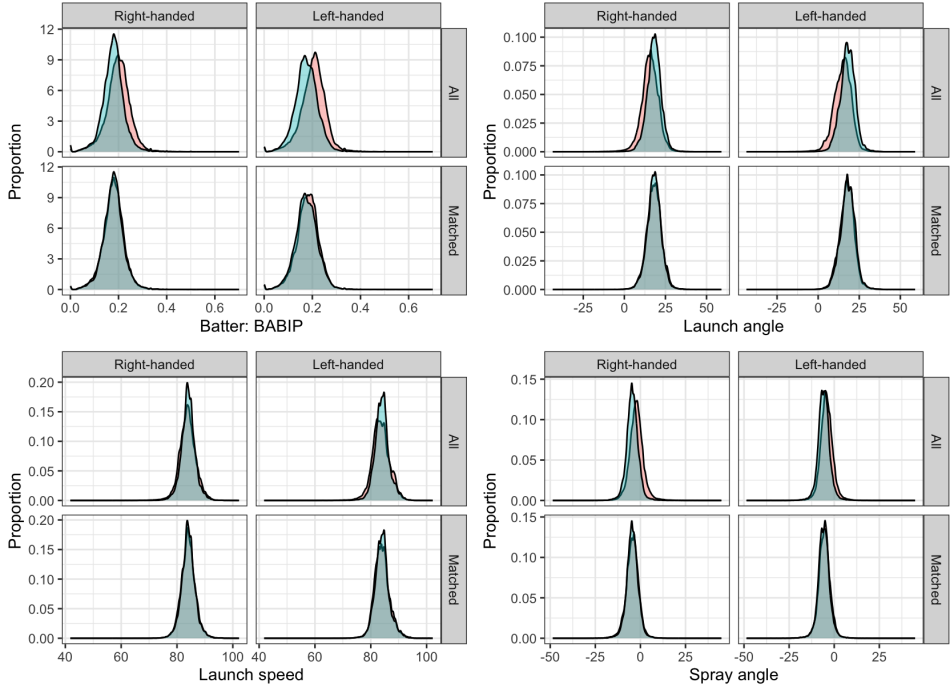}
\caption{Distributions of key variables in the original data set (All) and the chosen matched set (Matched), comparing shifted (blue) and non-shifted (red) plate appearances.}
\label{fig:match2}
\end{figure}

Balance in the matched set is improved over the original data set. Figure \ref{fig:match2} shows plots of the key variables: batter's BABIP and average hitting tendencies (launch speed, launch angle, and spray angle). The improvement in the balance from the original dataset to the matched is particularly noticeable for batter BABIP and launch angle. The distance between distributions of shifted and non-shifted plate appearances can be measured by the distance between eCDFs, considering both the mean distance between the eCDFs of a covariate across the groups and the maximum distance between the eCDFs. Amongst covariates, the largest difference in the distance between eCDFs was spray angle for right-handed batters (mean and max are 0.015 and 0.029) and launch speed for left-handed batters (mean and max are 0.015 and 0.033).

\subsection{Inverse probability of treatment weighting}
A possible limitation of $k:1$ nearest neighbour matching methods is that not all observations are used. A covariate balancing method that uses all the data is inverse probability of treatment weighting (IPTW), or in the case of estimating ETT, weighting by the odds \citep{Stuart2010}. Based on the Horovitz-Thompson estimator \citep{MostlyHarmlessEconometrics:AngristEtAl2008}, the IPTW estimator for ETT is

\begin{equation}\label{IPTW}
\mathbb{E}[Y|T=1]-\frac{1}{\mathbb{P}(T=1)}\mathbb{E}\left[(1-T)\frac{\mathbb{P}(T=1|\mathbf{C})}{\mathbb{P}(T=0|\mathbf{C})}Y\right]
\end{equation}

As with matching, we utilized a linear propensity score distance computed on the original (unmatched) dataset, $\widehat{\pi}(\mathbf{C})=\mathbb{P}(T=1|\mathbf{C})$. The estimate of the ETT was calculated using \eqref{IPTW}. We bootstrapped 10000 estimates to get a sampling distribution and took the standard deviation and quantiles (2.5\% and 97.5\%) to estimate the standard error and 95\% confidence intervals, respectively. Because of the form of the IPTW estimator, it can be susceptible to over-weighting a few observations when the propensity score is close to 1 \citep{Stuart2010}. An analysis of the impact of observations with high propensity scores is given in the Supplementary material.

\subsection{Instrumental variables}\label{sec:IVs}
We put forward the fielding team's propensity to shift as an instrument due to its similarity to what is known as a \textit{preference-based instrument} in health studies, so-called because the variable represents groups with different tendencies or preferences to treat. To be a valid IV, the fielding team's propensity to shift must satisfy the assumptions of relevance, effective random assignment, exclusions restriction, and no effect modification by the IV, as detailed in Section \ref{sec:methods}. Here, we explain why these assumptions may be reasonable for the fielding team's propensity to shift. We include a line of reasoning for the possibility that these assumptions are only valid under certain conditions, specifically, when conditioned on game year and pitcher measurements (including the pitcher's BABIP).

\subsubsection*{Relevance}
We posit that a fielding team's previous tendency to shift is correlated with the team's decision to shift during a given plate appearance. Specifically, a team with a history of shifting frequently is more likely to shift during a particular plate appearance. We performed a linear regression of fielding alignment on the fielding team's propensity to shift, or `treatment' on `instrument', yielded an $F$-statistic of $1.18 \times 10^{5}$, or a partial $F$-statistic of $5.74 \times 10^4$ when conditioned on game year and pitcher measurements. The $F$-statistics for subgroups defined by batter-handedness were of a similar order of magnitude: $8.26 \times 10^{4}$ (unconditioned) and $5.11\times 10^4$ (conditional) for right-handed batters, $5.15 \times 10^{4}$ (unconditioned) and $1.79 \times 10^4$ (conditional) for left-handed batters. All $F$-statistics are sufficiently large in both the unconditioned and conditional cases, indicating an association between the treatment assignment and the proposed instrument. This supports the assumption that the team's propensity to shift is a relevant instrument.

\subsubsection*{Effective random assignment}
To assess the validity of the assumption of effective random assignment, we examine the relationship between the fielding team's tendency to shift and the measured variables that may affect the treatment and outcome. If the assumption of effective random assignment holds, the distribution of a variable should be consistent across each level of the independent variable unless that variable is controlled for. While the assumption is untestable for unmeasured variables, we can verify all measured variables. Doing so serves two purposes: it helps support the selection of variables controlled for in the IV analysis and highlights variables that could be proxies for unmeasured common causes. In the latter scenario, the assumption would not hold.

\begin{figure}[!ht]
\centering
\includegraphics[width=0.9\linewidth]{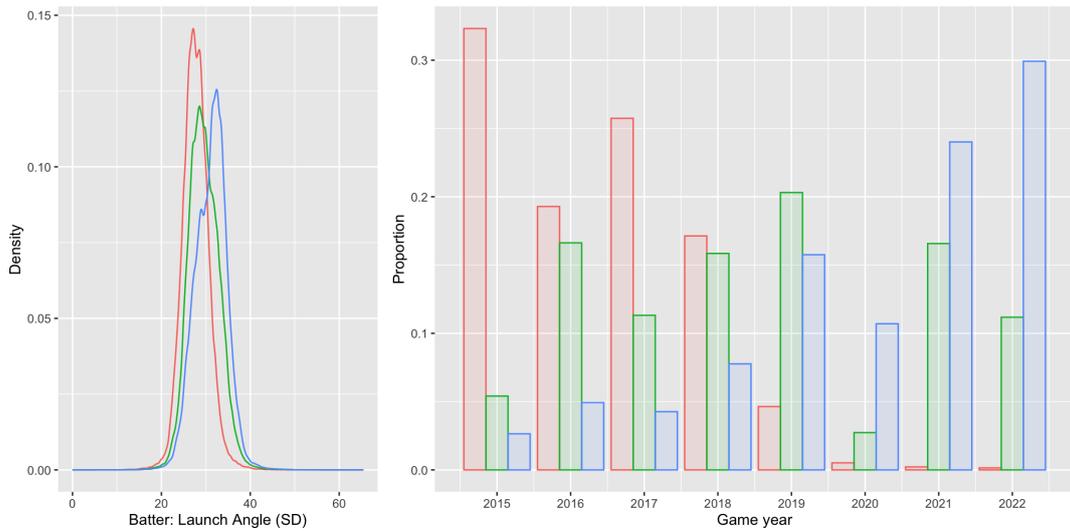}
\caption{The distributions of the variables, launch angle (SD) and game year, vary with the proposed instrument. Here, the instrument is categorized as low (red), medium (green), and high (blue) propensity to shift.}
\label{fig:iv1}
\end{figure}

While most variables did not show an association with the proposed instrument, two variables did: the variability of the batter's launch angle and the game year. Figure \ref{fig:iv1} shows the distribution of these two variables, grouped by the level of the team's shifting propensity---low, medium, high---as determined by quantile. Similar plots of all batter, pitcher, and context variables, split out by batter handedness, are included in the Supplementary Material.

Variability in the batter's launch angle could be considered a proxy for when a game is played in the season. As a season progresses, it is reasonable to assume that the variability of a batter's launch angle will decrease, and teams will settle into their defensive strategies. Thus, the length of time into a season could be a common cause of the instrument and this measured variable. However, the time into the season is not likely to affect the change in expected runs.

On the other hand, the game year could impact the change in expected runs, as the run expectancy matrix is specific to the year \citep{RE24:Weinberg2014} (as mentioned in Section \ref{sec:data}). Therefore, controlling for the game year may be necessary to satisfy the assumption of effective random assignment. While the main model will assume game year does not impact effective random assignment, we also considered an IV model that does control for game year. We allowed for interaction between the instrument--a team's propensity to shift--and a covariate--specifically, the game year--using the method from \citet{Wang+TchetgenTchetgen2018}. This method does not require the assumption that there is no effect modification by the IV.

\subsubsection*{Exclusion restriction}
\begin{figure}[!ht]
\centering
\includegraphics[width=0.3\linewidth]{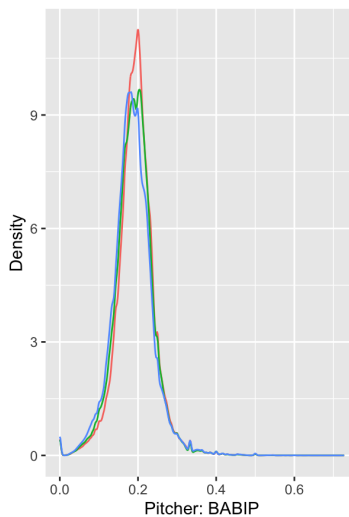}
\caption{The distribution of the pitcher's BABIP does not vary with the proposed instrument, shown as low (red), medium (green), and high (blue) propensity to shift.}
\label{fig:iv2}
\end{figure}

It is possible that a team's fielding abilities, whether good or bad, could be related to the team's shift rate. If this were the case, it would violate the exclusion restriction as the instrument would affect the outcome not solely through the treatment. However, the pitcher's BABIP partly reflects the quality of the fielders behind the pitcher. Figure \ref{fig:iv2} does not provide evidence that a team's shift rate is associated with the pitcher's BABIP. Therefore, we proceed with the assumption of exclusion restriction.

\subsubsection*{No effect modification by the instrument} \label{assume:noeffectmod}
Assuming there is no effect modification by the instrument means that all teams are equally effective when they shift, regardless of whether they use it more or less frequently. For example, a team that employs the shift more often is not necessarily doing so because they are better (or worse) at fielding in a shifted alignment than in a standard alignment. However, this assumption may not be plausible without controlling for the game year, as teams may make significant strategies in the off-season. For example, teams with high shift rates in 2022 may have been more (or less) effective when deploying the shift than teams with high shift rates in 2015. In controlling for the game year, we allowed for interaction between the instrument--a team's propensity to shift--and the covariate--the game year so as to avoid any unnecessary assumptions. We calculated the two-stage least squares estimate for each year and then obtained the estimate for the ETT by taking the weighted average, with each year's average weighted by its proportion of treated observations \citep{Wang+TchetgenTchetgen2018}.

\section{Results} \label{sec:results}
\begin{figure}[!ht]
\centering
\includegraphics[width=0.5\linewidth]{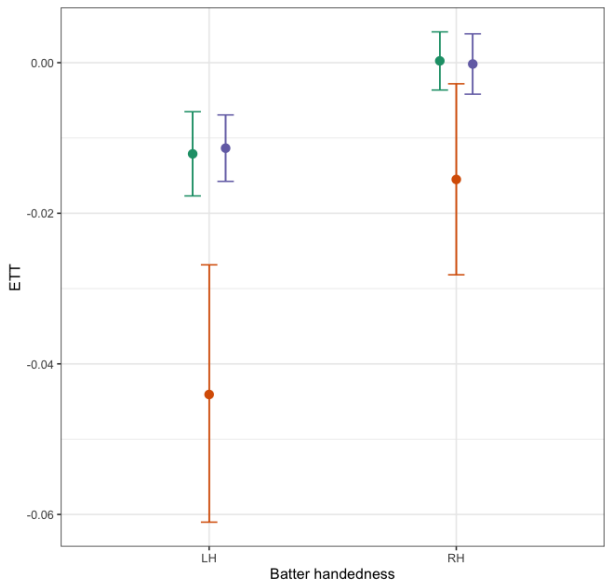}
\caption{Estimates (dots) and 95\% confidence intervals (bars) from all three methods (IPTW (green), IV (red), and matching (purple)).}
\label{fig:results3}
\end{figure}

To estimate the ETT on the matched set, we modelled the outcome with a linear regression model and employed g-computation \citep{R-marginaleffects} with robust standard errors. To estimate the ETT with IPTW, statistics were computed from the bootstrapped sampling distribution. To estimate the ETT using the IV, we ran a two-stage least-squares regression \citep{R-ivtools} in each year, taking a weighted average amongst shifted plated appearance to get the marginal effect; confidence intervals were constructed using the bootstrap. The results for all methods are shown in Figure \ref{fig:results3}.

Confidence intervals for both confounder balancing methods---matching and IPTW---are mostly overlapping, although the IPTW interval for left-handed batters is wider. The IV estimates are lower than the balancing estimates for both right-handed and left-handed batters, and with wider confidence intervals. Wider confidence intervals for the IV are not unexpected since two-stage least squares can be susceptible to the propagation of errors, and the assumptions of linearity needed may lead to model misspecification, a topic which we will return to in Section \ref{sec:discussion}. All three methods showed that the infield shift was effective when used against left-handed batters, as the upper bounds of the confidence intervals were less than zero. Similarly, point estimates of the ETT for left-handed batters were lower than those for right-handed batters in all methods. However, only the IV model suggested that the defensive impact of the shift was non-zero when used against right-handed batters. The intervals from the balancing methods indicate moderation of the effect of the shift by batter-handedness, whereas the IV results leave open the possibility that batter-handedness did not change the effect of the shift since there is a small overlap between those confidence intervals.  


\section{Discussion} \label{sec:discussion}
Causal conclusions from observational studies depend on various assumptions that can be justified but which are not testable from the data. The risk of making untestable assumptions is that if they are incorrect, then the causal conclusions are invalid. However, this risk can be mitigated by analyzing the question using different sets of assumptions. In this analysis, we used balancing methods---matching and IPTW---which rely on the assumption that all confounding is measured, complimented by the IV method, which controls for unmeasured confounding by relying on different assumptions (as discussed in Section \ref{sec:methods}). While estimates obtained by the various methods differed, some conclusions were similar: the shift is likely effective against left-handed batters. Only the IV model suggested the shift may be effective for all batters. 

Both sets of methods are vulnerable to model misspecification. In particular, balancing methods rely on correctly specifying the propensity score model, and the two-stage least squares estimator for the IV estimand requires linearity of both the treatment and outcome models. Doubly robust estimators require that only one of the treatment or outcome models be correctly specified, and thus, estimating the infield shift with a doubly robust estimator may be an avenue for future research. 

Our estimand is based on Statcast's empirical categorization of infielder positioning into shifted versus non-shifted plate appearances in MLB from 2015 to 2022. Baseball teams may use different internal definitions of the shift, even having multiple versions, and the effectiveness of these variations could be assessed using similar approaches as in this analysis, but utilizing proprietary data. Additionally, this analysis excluded batters without hit trajectory data and pitchers without pitch trajectory data over the course of the season. Thus, these results only apply to similar situations, specifically, to plate appearances where the batter and pitcher will record trajectory data, hit or pitch, respectively, during the season. A possible extension of the analysis here is one that includes imputation of the missing data. 

The results of this study indicate that the infield shift has been an effective defensive strategy against targeted left-handed batters. To contextualize the effect of the shift, we can estimate the magnitude of the shift's impact on a game. In a standard nine-inning baseball game, each team has at least 27 plate appearances, which is of the order $10^1$. Our most conservative estimates indicate that the shift has an approximate effect of $10^{-2}$ on targeted left-handed batters, with potentially no effect on targeted right-handed batters. Left-handed batters represent approximately $35\%-39\%$ of a team's plate appearances, and at its peak, the infield shift was used in $55\%$ of these plate appearances. This suggests that, at most, the infield shift may be causing a decrease of approximately $10^{-1}$ expected runs per game. Although this may confer a competitive advantage between teams over the course of a season, the average spectator might not readily discern this disparity. We note that this estimated effect is based on the ETT, and thus, does not encompass potential effects on plate appearances not subject to the shift.

The rule change impacts the plate appearances that would have been shifted as well as those that would not have been targeted with the shift. As such, our estimates do not capture the total impact of the rule change. To ascertain whether the rule change will be perceptible, a different causal estimand, such as the average treatment effect (ATE), would be necessary to represent the effect of the infield shift on all plate appearances. Moreover, a spectator's experience may be less dependent on the runs scored and more influenced by the overall excitement in a game. Whether the infield shift suppresses the exciting aspects of a game is a separate question from the one explored in this analysis. Nevertheless, investigating the effect of the infield shift on baseball's dynamism could be undertaken using the same methodologies presented here, albeit with a different outcome variable, such as the number of balls put into play. Estimating the ATE and exploring alternative outcome variables are potential avenues for future research.



\bibliographystyle{agsm} 

\newpage

\appendix

\renewcommand\thefigure{\thesection.\arabic{figure}}    

\setcounter{figure}{0}    

\section{Supplementary material}

\subsection{Variables}

The full list of continuous variables used, along with summary statistics, are shown in Tab. \ref{tab:sumstats}. Counts for discrete variables are given in Tab. \ref{tab:counts}. The Statcast fields are all defined in \citet{statcast_dd} and we described how we calculated the variables in Section \ref{sec:data}. Note that all batter and pitcher variables were used as confounders, as well as game year and the pitcher's handedness, while the batter's handedness defined subgroups for the analysis. The team's propensity to shift was used as an instrumental variable. The shifted fielding alignment was considered as the treatment variable and the change in run expectancy was the outcome variable. 

\begin{table}[hbt]
\centering
\caption{Summary statistics for continuous variables}
\label{tab:sumstats}
\begin{tabular}{l RRRRR}
\toprule
Variable & \text{Mean} & \text{SD} & \text{Median} & \text{Min} & \text{Max}\\
\midrule
Batter: BABIP & 0.19 & 0.05 & 0.20 & 0.00 & 0.70\\
Batter: BABIP (SD) & 0.39 & 0.05 & 0.40 & 0.00 & 0.53\\
Batter: Launch angle (SD) & 29.46 & 3.61 & 29.24 & 0.00 & 65.42\\
Batter: Launch angle (avg) & 15.90 & 5.23 & 16.15 & -43.57 & 59.00\\
Batter: Launch speed (SD) & 14.57 & 1.68 & 14.55 & 0.07 & 39.88\\
Batter: Launch speed (avg) & 83.79 & 3.00 & 83.78 & 41.80 & 102.20\\
Batter: Spray angle (SD) & 21.19 & 1.88 & 21.15 & 0.42 & 68.26\\
Batter: Spray angle (avg) & -3.33 & 3.95 & -3.45 & -48.60 & 44.85\\
\addlinespace
Batter: Strikeout (SD) & 0.41 & 0.05 & 0.41 & 0.00 & 0.53\\
Batter: Strikeout (avg) & 0.22 & 0.08 & 0.21 & 0.00 & 0.86\\
Batter: Walk (SD) & 0.28 & 0.07 & 0.29 & 0.00 & 0.52\\
Batter: Walk (avg) & 0.09 & 0.04 & 0.09 & 0.00 & 0.64\\
Batter: wOBA & 0.33 & 0.07 & 0.33 & 0.00 & 1.07\\
Batter: wOBA (SD) & 0.51 & 0.07 & 0.52 & 0.00 & 1.05\\
\addlinespace
Pitcher: BABIP & 0.19 & 0.05 & 0.19 & 0.00 & 0.73\\
Pitcher: BABIP (SD) & 0.39 & 0.05 & 0.39 & 0.00 & 0.53\\
Pitcher: Plate x (SD) & 0.84 & 0.08 & 0.84 & 0.29 & 1.75\\
Pitcher: Plate x (avg) & 0.04 & 0.16 & 0.04 & -0.88 & 0.88\\
Pitcher: Plate z (SD) & 0.93 & 0.11 & 0.92 & 0.40 & 1.61\\
Pitcher: Plate z (avg) & 2.25 & 0.19 & 2.25 & 1.40 & 3.81\\
Pitcher: Release speed (SD) & 5.10 & 1.07 & 5.06 & 0.68 & 19.80\\
Pitcher: Release speed (avg) & 88.75 & 2.89 & 88.83 & 51.19 & 99.17\\
\addlinespace
Pitcher: Spin rate (SD) & 258.14 & 98.56 & 247.08 & 9.87 & 877.65\\
Pitcher: Spin rate (avg) & 2218.85 & 183.90 & 2212.66 & 559.28 & 3055.86\\
Pitcher: Strikeout (SD) & 0.41 & 0.05 & 0.42 & 0.00 & 0.53\\
Pitcher: Strikeout (avg) & 0.23 & 0.07 & 0.22 & 0.00 & 0.82\\
Pitcher: Walk (SD) & 0.28 & 0.06 & 0.28 & 0.00 & 0.53\\
Pitcher: Walk (avg) & 0.09 & 0.04 & 0.09 & 0.00 & 0.70\\
Pitcher: wOBA & 0.32 & 0.07 & 0.32 & 0.00 & 1.04\\
Pitcher: wOBA (SD) & 0.51 & 0.06 & 0.51 & 0.00 & 0.97\\
\addlinespace
Team: Propensity to shift & 0.22 & 0.13 & 0.20 & 0.00 & 0.92\\
\addlinespace
Change in run expectancy & 0.00 & 0.48 & -0.17 & -1.41 & 3.38\\
\bottomrule
\end{tabular}
\end{table}

\begin{table}[htb]
\centering
\caption{Counts by year and binary variables}
\label{tab:counts}
\begin{tabular}{lllRRRRRRRR}
\toprule
Batter & Pitcher & Shifted & \text{2015} & \text{2016} & \text{2017} & \text{2018} & \text{2019} & \text{2020} & \text{2021} & \text{2022}\\
\midrule
RH & RH & No & 60949 & 62505 & 64794 & 57506 & 54684 & 16730 & 51925 & 53652\\
RH & RH & Yes & 3203 & 5406 & 4999 & 7683 & 11612 & 5467 & 11762 & 14509\\
RH & LH & No & 24589 & 24143 & 24643 & 25711 & 22558 & 6505 & 23513 & 21700\\
RH & LH & Yes & 1219 & 1957 & 1314 & 3040 & 5292 & 2775 & 6606 & 6785\\
\addlinespace
LH & RH & No & 34140 & 30669 & 30868 & 26633 & 20721 & 6222 & 17040 & 17645\\
LH & RH & Yes & 8975 & 11273 & 10914 & 13773 & 19459 & 8323 & 23018 & 23251\\
LH & LH & No & 11013 & 9541 & 9792 & 10134 & 8613 & 2396 & 7482 & 6505\\
LH & LH & Yes & 2383 & 2688 & 2622 & 3363 & 4737 & 2189 & 6706 & 5614\\
\bottomrule
\end{tabular}
\end{table}

\subsection{Matched sets}

Two sets of parameter choices emerged as possible options for analysis:
\begin{enumerate}[label=(\Alph*)]
    \item Ratio: 2:1, Re-use maximum = 4, Caliper = 0.20
    
    \item Ratio: 3:1, Re-use maximum = 5, Caliper = 0.15
\end{enumerate}

\begin{figure}[!hbt]
\centering
\includegraphics[width=0.8\linewidth]{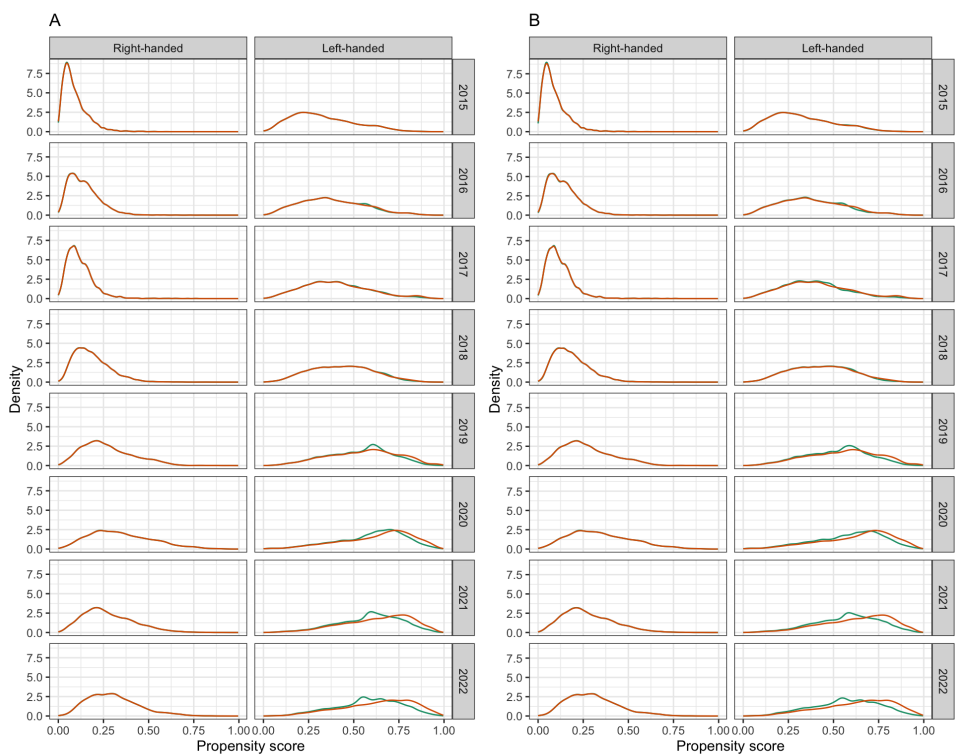}
\caption{The propensity score densities in the original dataset are shown in red, with the matched set propensity score densities shown in green for each option set of parameters.}
\label{fig:matchbalance1}
\end{figure}

Option B includes more unique observations than Option A and has a greater effective sample size. Both options have common support, even in subgroups defined by batter handedness and game year, as shown in Figure \ref{fig:matchbalance1}.

\begin{figure}[!htb]
\centering
\includegraphics[width=0.9\linewidth]{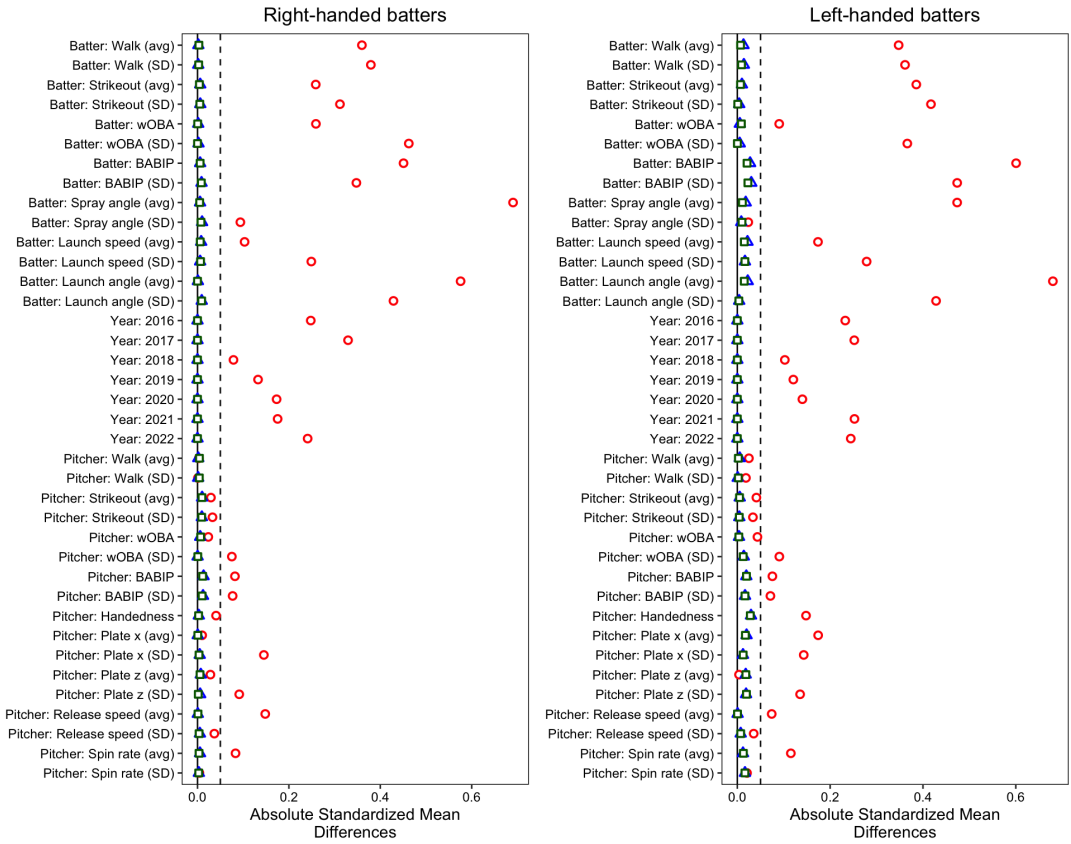}
\caption{Comparison of covariate balance between the full dataset (red circles), Option A (blue triangles), and Option B (green squares). The dotted line indicates the 0.05 threshold in units of the standardized mean difference.}
\label{fig:matchbalance2}
\end{figure}

Figure \ref{fig:matchbalance2} shows that both options improve on the balance in the full dataset. Both are well-balanced for all variables, with no variables having an absolute standardized mean difference outside a threshold of 0.05 (under 0.25 is recommended \citep{Stuart2010}). Although under the threshold, there is less balance amongst left-handed batters in a few variables. Amongst variables where there is less balance for left-handed batters, average launch angle, and the cumulative mean and standard deviation of batter BABIP stand out as those that could have more impact on the outcome, and Option B has a better balance for these variables (0.0171, -0.0197, -0.0208, respectively) than Option A (0.0253, -0.0258, -0.0274). Since an exact match was required for the game year, variables representing the game year are entirely balanced in both matched options.

Both options have good balance and common support. The clearest difference between the matched sets is in sample size, and thus, we chose to perform the analysis on Option B.

\subsection{Sensitivity of IPTW estimates}

Propensity scores close to 1 can have an outsized impact on IPTW estimates due to the form of the estimand. To check that a few observations are not skewing our estimate, we removed observations with the top 5\% of propensity scores and repeated the estimation procedure. The results in Table \ref{tab:IPTWsensit} show little difference when the observations with high propensity scores are removed.


\begin{table}[htb]
    \centering
    \caption{IPTW estimates are similar between the two datasets.}
    \begin{tabular}{rllrcrcr}
          & & && Estimate && $95\%$ confidence interval & \\
          \hline \\
         & RH & && && & \\
         & & All data && $0.00024$ && $(-0.00363, 0.00409)$ & \\
         & & Partial data$^*$ && $0.00028$ && $(-0.00319, 0.00379)$ &\\
         \hline \\
         & LH & && && & \\
         & & All data && $-0.01211$ && $(-0.01770, -0.00651)$ & \\
         & & Partial data$^*$ && $-0.00990$ && $(-0.01410, -0.00572)$ &\\
        \multicolumn{8}{c}{\footnotesize $^*$Observations with propensity scores up to the $95^{th}$ percentile.}
    \end{tabular}
    \label{tab:IPTWsensit}
\end{table}

\newpage

\end{document}